\documentclass[twocolumn, preprintnumbers, amsmath, amssym]{revtex4-1}
\usepackage{graphicx}
\usepackage{dcolumn}
\usepackage{bm}
\usepackage{enumitem}
\usepackage{amssymb}
\usepackage{amsmath}
\usepackage{mathrsfs}
\usepackage{array}
\usepackage{setspace}
\usepackage{multirow}
\usepackage{float}
\begin{document}
\newcolumntype{C}[1]{>{\centering\arraybackslash}p{#1}}


\title{Measurement of enhanced electric dipole transition strengths at high spin in $^{100}$Ru - possible observation of
octupole deformation}
	
	\author{A. Karmakar$^{1,2}$\thanks{\email anindita.karmakar@saha.ac.in},
    Nazira Nazir$^{3}$ \thanks{\email naziranazir238@gmail.com},
		P. Datta$^{4}$,\thanks,
		J. A. Sheikh,$^{3,5}$,
		S. Jehangir$^{5}$,
            G. H. Bhat$^{6}$,
		S. S. Nayak$^{2,7}$,
		Soumik Bhattacharya$^{2,7}$,
		Suchorita Paul$^{2,7}$,
		Snigdha Pal$^{2,7}$,
		S. Bhattacharyya$^{2,7}$,
		G. Mukherjee$^{2,7}$,
		S. Basu$^{2,7}$,
		S. Chakraborty$^{2,7}$,
		S. Panwar$^{2,7}$,
		Pankaj K. Giri$^{8}$,
		R. Raut$^8$,
		S. S. Ghugre$^8$,
		R. Palit $^9$,
		Sajad Ali$^{10}$,
		W. Shaikh$^{11}$
	}
	\author{S. Chattopadhyay$^{1,2}$ \thanks{\email sukalyan.chattopadhyay@saha.ac.in}}
	
	\affiliation{$^{1}$Saha Institute of Nuclear Physics, 1/AF, Bidhan Nagar, Kolkata 700064, India}
	\affiliation{$^{2}$Homi Bhabha National Institute, Training School Complex, Anushakti Nagar, Mumbai - 400094, India}
        \affiliation{$^{3}$Department of Physics, University of Kashmir, Srinagar, 190 006, India}
	\affiliation{$^{4}$Ananda Mohan College, Kolkata- 700009, India.}
	\affiliation{$^{5}$Department of Physics, Islamic University of Science and Technology, Awantipora, 192 122, India}
       \affiliation{$^{6}$Department of Higher Education (GDC Shopian), Jammu and Kashmir, 192 303, India}
        
	\affiliation{$^{7}$Variable Energy Cyclotron Centre, Kolkata - 700064, India}
	\affiliation{$^{8}$UGC-DAE Consortium for Scientific Research, Kolkata Centre, Kolkata 700106, India}
	\affiliation{$^{9}$Tata Institute of Fundamental Research, Mumbai- 400005, India}
	\affiliation{$^{10}$Government General Degree College at Pedong, Kalimpong 734311, India}
	\affiliation{$^{11}$Mugberia Gangadhar Mahavidyalaya, Purba Medinipur, India.}
	
	\date{\today}

        \begin{abstract}
The majority of atomic nuclei have deformed shapes and nearly all these shapes are symmetric with respect to reflection. There are only a few reflection asymmetric pear-shaped nuclei that have been found in actinide and lanthanide regions, which have static octupole deformation. These nuclei possess an intrinsic electric dipole moment due to the shift between the center of charge and the center of mass. This manifests in the enhancement of the electric dipole transition rates. In this article, we report on the measurement of the lifetimes of the high spin levels of the two alternate parity bands in $^{100}$Ru through the Doppler Shift Attenuation Method. The estimated electric dipole transition rates have been compared with the calculated transition rates using the triaxial projected shell model without octupole deformation, and are found to be an order of magnitude enhanced. Thus, the observation of seven inter-leaved electric dipole transitions with enhanced rates establish $^{100}$Ru as possibly the first octupole deformed nucleus reported in the $A \approx 100$ mass region.\end{abstract}

\maketitle

The breakdown of the reflection symmetry is known to modify the properties of bulk materials as well as the quantum systems. The polar acentric
crystal classes break the point reflection symmetry. This symmetry states that if there is an atom at (x, y, z) relative to the center of symmetry, there must also exist an atom at (-x, -y, -z) and this is true for all x, y, z. These polar crystals possess a dipole moment and exhibit technologically important properties like ferroelectricity and pyroelectricity \cite{coulson_1958}. In contrast, the reflection symmetry in a quantum system is broken due to the shape anisotropy. For example, the ZnO nano-prisms possess an electric dipole moment, which influences the band structure thereby modifying their optical properties \cite{article2}. The other well-studied quantum system of this class is the pear-shaped atomic nucleus, which also possesses an intrinsic dipole moment. It arises due to the separation between the center of mass and the center of charge as the concentration of protons is higher in the region of higher curvature, which is the narrower end of the pear \cite{BM}. In recent years, these nuclei have attracted considerable experimental attention as the atoms with pear-shaped nuclei are ideal candidates for the search for permanent atomic electric dipole moment, which is indicative of CP violation and physics beyond the Standard Model \cite{gaffney, Chishti:2019emu, Butler2019}.

The breakdown of reflection symmetry and the presence of an intrinsic dipole moment leads to a characteristic rotational band structure for an octupole-deformed even-even nucleus, where two alternating parity bands are observed connected by relatively fast electric dipole ($E1$) transitions. The even-even isotopes of Ra – Th \cite{Chishti:2019emu, gaffney, PhysRevLett.78.2920, PhysRevLett.124.042503} ($Z \approx 88$ and $N \approx 134$) and Sm–Ba ($Z \approx 56$ and $N \approx 88$) nuclei \cite{PhysRevLett.124.032501,ba2,ba3,ba4} exhibit these properties and are examples of nuclei with permanent octupole deformation. In a recent communication, \cite{100ru_us}, seven inter-leaved electric dipole ($E1$) transitions have been observed in $^{100}$Ru between the alternate parity bands. However, the estimated $B(E1)/B(E2)$ values were found to be small compared to those reported for well-established octupole-deformed nuclei. Thus, to establish octupole collectivity in $^{100}$Ru, a direct level lifetime measurement is essential to estimate the $B(E1)$ rates and compare them with the calculated single particle transition rates.\par

To perform this measurement, we have populated the excited levels of $^{100}$Ru through $^{100}$Mo($^{4}$He,4n)$^{100}$Ru reaction at a beam energy of 45 MeV delivered by the K - 130 cyclotron at the Variable Energy Cyclotron Centre, Kolkata. The $\gamma$ rays were detected using the Indian National Gamma Array (INGA) \cite{inga} consisted of 11 Compton-suppressed clover detectors placed at 40$^\circ$ (two detectors), 90$^\circ$ (six detectors) and 125$^\circ$ (three detectors) with respect to the beam direction. The target was a 10 mg/cm$^2$ thick foil of $^{100}$Mo. The time-stamped data were recorded by a digital data acquisition system based on a 250 MHz, 12-bit PIXIE - 16 digitizer (XIA LLC) \cite{DAS2018138}. A total of 4 $\times$ 10$^9$ $\gamma$-$\gamma$ coincidence events were sorted within a time window of 100 ns by using the sorting program BINDAS \cite{bindas}. About 58\% events were found to belong to the $^{100}$Ru channel. The data were sorted to form angle-dependent asymmetric matrices with 90$^\circ$ detectors on one axis and forward or backward detectors on the other axis. The lineshapes are extracted using 1023 keV (10$^+$ $\xrightarrow{}$ 8$^+$) for Band 2 and 727 keV (11$^-$ $\xrightarrow{}$ 9$^-$) for Band 3. The nomenclatures for the band structures are defined in Ref.~\cite{100ru_us}. The lineshape analyses were carried out using the LINESHAPE package \cite{lineshape} along with the development reported in Ref.~\cite{raj_dsam}. The velocity profiles for the $^{100}$Ru residues at the three angles of 40$^\circ$, 90$^\circ$, and 125$^\circ$ were simulated using the stopping powers calculated by SRIM \cite{srim}. The latter is merited with lesser (5\%) uncertainties \cite{raj_dsam} vis-a-vis the stopping power modeled by one of the algorithms (Ziegler or Northchiffe - Schilling \cite{schi}) in the original lineshape package. These profiles were calculated in time steps of 0.001 ps for 10000 trajectories while considering their origins distributed across the expanse of the thick self-supporting, and entirely productive, target. It may be noted that only an insignificant ($\approx$ 4\%) proportion of trajectories remain unfinished within the target, that do not impact the subsequent analysis. The $\gamma$-ray energies and intensities were treated as the inputs to the line shape fit. The side-feeding intensities at each level were calculated from the measured intensities reported in Ref.~\cite{100ru_us}. The side-feeding to each level has been modeled as a cascade of five transitions with a moment of inertia which is comparable to the band of interest. The quadrupole moments of the side-feeding sequence were allowed to vary, which, when combined with the moment of inertia, gave an effective side-feeding time for each level \cite{104pd, 106ag}.\par

\begin{figure}[!ht]
    \begin{center}
       \hspace*{-0.2cm}\includegraphics[height=7.5cm, width=9cm, angle =0]{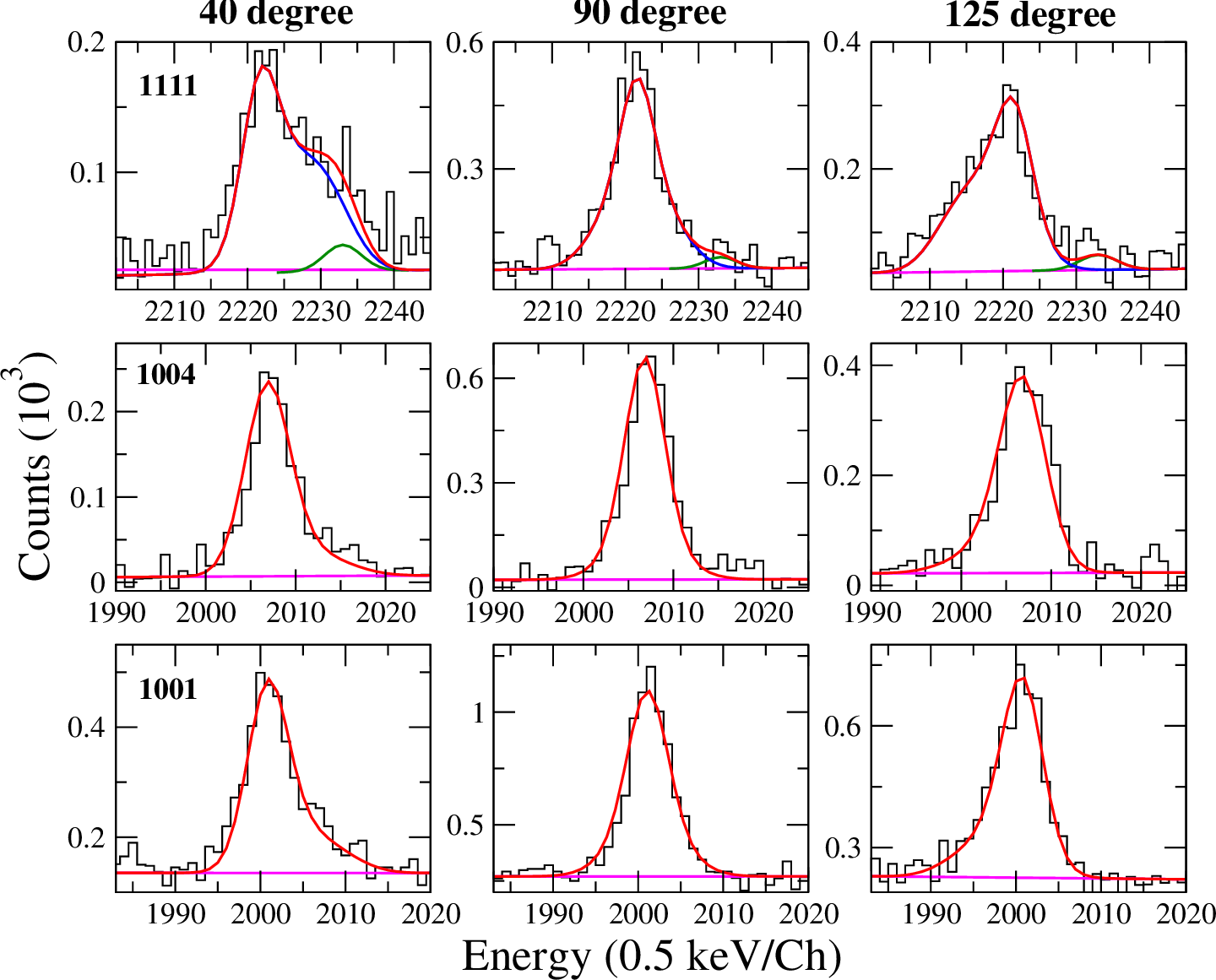}
    \end{center}
    \caption{(Color online) The line shape fits for 1001 and 1111 keV of Band 2 and 1004 keV of Band 3 at 40$^\circ$, 90$^\circ$ and 157$^\circ$ to the beam direction. The fitted Doppler broadened line shapes are drawn in blue lines while the contaminant peaks are shown in green lines. The result of the fit to the experimental data is shown in red lines.}
    \label{shape}
\end{figure}

The global fit for each band was carried out simultaneously at 40$^\circ$, 90$^\circ$, and 125$^\circ$. The lineshape fits for three $\gamma$ transitions, namely 1001 keV (16$^+$ $\xrightarrow{}$ 14$^+$ for Band 2), 1004 keV (17$^-$ $\xrightarrow{}$ 15$^-$ for Band 3) and 1111 keV (18$^+$ $\xrightarrow{}$ 16$^+$ for Band 2) are shown in Fig.~\ref{shape}. The measured level lifetimes and the evaluated transition rates are listed in Table.~\ref{table1}. The quoted uncertainties on the level lifetimes were calculated by adding in quadrature the uncertainties due to the fitting and side feeding intensity \cite{104pd, 106ag, 110ag}.

The modeling of the side-feeding lifetimes was cross-checked by fitting the observed line shapes of 795 (14$^+$ $\xrightarrow{}$ 12$^+$) and 932 (13$^-$ $\xrightarrow{}$ 11$^-$) keV transitions extracted from the top gates of 1001 (16$^+$ $\xrightarrow{}$ 14$^+$) and 1036 (15$^-$ $\xrightarrow{}$ 13$^-$) keV, respectively. In these fits, only the effective lifetimes of the 16$^+$ and 15$^-$ levels were considered. The estimated lifetimes for the 14$^+$ and 13$^-$ levels were found to be 1.07(5) and 0.44(3) ps respectively, which match with the values reported in Table.~\ref{table1} within $\pm$ 1$\sigma$. A similar exercise for the higher spin levels was not possible due to inadequate statistics.

\begin{table}
\caption{The measured lifetimes of electric quadrupole transitions of the $^{100}$Ru levels for Band 2 and Band 3.
The uncertainties in the lifetime measurements were derived from the behavior of the $\chi^2$ in the vicinity of the minimum for the simultaneous fit at the three angles. The $E1$ and $E2$ branching ratios for each level are adapted from the intensity table of Ref.~\cite{100ru_us}.}
\label{table1}
\vspace{1mm}
\small
\begin{center}
\begin{tabular}{C{1.5cm} C{1.3cm} C{1.3cm} C{1.3cm} C{1.5cm}}
\hline
Fitting $E_\gamma$ & $J_\pi^i$ & $\tau$ & $B(E2)$ & $B(E1)$ \\
(keV) & & (ps) & (e$^2$fm$^4$) & ($\times$ 10$^{-4}$ e$^2$fm$^2$)\\
\hline
& & \textbf{Band 2} & & \\
\hline
795.3(3) & 14$^+$ & 1.02(9) & 2418(214) & 1.4(4) \\
1000.5(3) & 16$^+$ & 0.32(3) & 2465(232) &2.4(6) \\
1111.4(4) & 18$^+$ & 0.17(1) & 2672(169) &7.7(25) \\
1229.5(5) & 20$^+$ & $<$ 0.14$^u$ & $>$ 2074$^l$ &- \\
\hline
& & \textbf{Band 3} & & \\
\hline
931.9(2) & 13$^-$ & 0.40(3) & 1967(148) & 2.0(4)\\
1035.7(3) & 15$^-$ & 0.31(2) & 2139(139)&2.7(6)\\
1003.5(4) & 17$^-$ & 0.29(2) & 2672(189)&4.4(15)\\
1253.7(5) & 19$^-$ & $<$ 0.12$^u$ & $>$ 2194$^l$ & -\\
\hline
\end{tabular}
\end{center}
\vspace{-0.5cm}
\begin{itemize}[label={}]
\setlength{\itemsep}{0.5pt}
    \item $^u$: upper limit (effective lifetime)
    \item $^l$: lower limit
\end{itemize}
\end{table}


\begin{figure}[!ht]
       \hspace*{0cm}\includegraphics[height=10.5cm, width=8cm, angle =0]{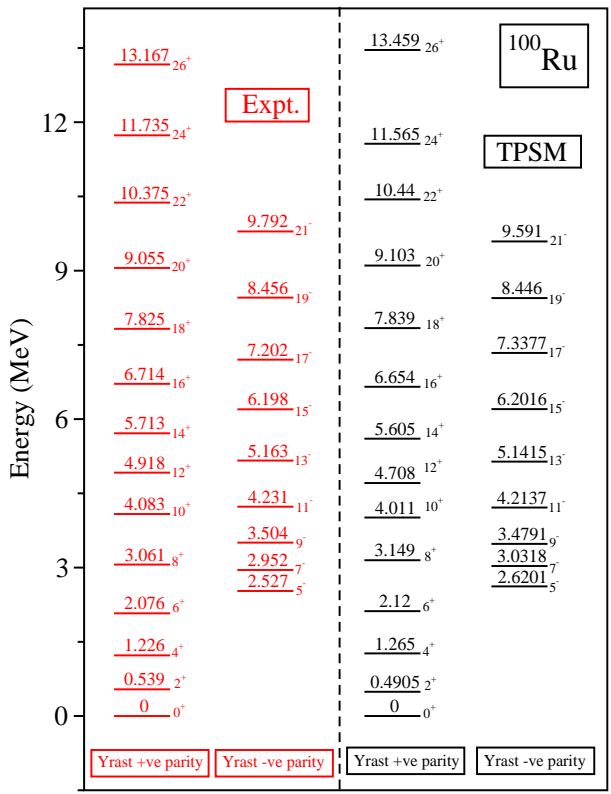}
    \caption{(Color online) TPSM projected energies after configuration mixing are compared with experimental data \cite{100ru_us} for $+$ve and $-$ve parity bands in $^{100}$Ru.}
    \label{level_energy}
\end{figure}

The extracted $B(E2)$ values from the level lifetimes are nearly the same for Band 2 and 3 as is evident from Table.~\ref{table1}. The weighted average of these values leads to an average quadrupole moment ($Q_0$) of 264(8) efm$^2$ for these bands. It may be observed from the table that the $B(E1)$ values remain nearly constant till $I = 16\hbar$. The weighted mean of these four values leads to a $D_0$ value of 0.04(1) efm. Beyond this spin, there is a clear indication of the enhancement of $B(E1)$ rates which correspond to a weighted average $D_0$ value of 0.07(2) efm. In this connection, it may be noted that the parity splitting between Band 2 and Band 3 also vanishes in this spin domain. Similar observations in the lanthanide region have been associated with the emergence of stable octupole deformation \cite{finger}.\par
The single particle $B(E1)$ rate of 1.8(4) $\times$ 10$^{-6}$ e$^2$fm$^2$ was measured for the 10$^+$ [h$_{11/2}^2$] $\xrightarrow{}$ 9$^-$ $[h_{11/2}$ $\otimes$ $\{g_{7/2}/d_{5/2}\}]$ transition in the neighboring $^{110}$Cd \cite{110cd}. Thus, the observed $B(E1)$ rates in $^{100}$Ru are two orders of magnitude enhanced compared to that observed rate in $^{110}$Cd \cite{110cd}. This is indicative of the presence of octupole collectivity in $^{100}$Ru. However, the $E1$ transition rates are dominated by single particle contributions, which can be different for $^{110}$Cd and $^{100}$Ru. Thus, it is necessary to estimate these rates for $^{100}$Ru.\par

To elucidate the observed properties of $^{100}$Ru, we have extended the triaxial projected shell model (TPSM) approach to include octupole-octupole interaction in the shell model Hamiltonian. In the original version of TPSM, the Hamiltonian consists of pairing and quadrupole-quadrupole interaction terms and is designed to describe the quadrupole properties of atomic nuclei \cite{JAS99}. The validity and applicability of the TPSM approach to describe a broad range of the observed properties in deformed and transitional nuclei is now well established \cite{JEH18, JEH21, EPJ21, JEH22, NAZ23,spr24}. The TPSM approach has been recently generalized to investigate the negative parity band structures in even-even systems \cite{mus21, NAZ_23}. In the earlier version of the TPSM approach, the valence particles were restricted to occupy a single oscillator shell, and it was possible to investigate only $+$ve parity states. In the recent extension \cite{NAZ_23}, the valence particles are allowed to occupy two major oscillator shells, and this allows to construct the $-$ve parity states in an even-even system.\par

The transitions between the $-$ve and $+$ve parity states are quite weak, as will be discussed in the following, using the pairing and quadrupole-quadrupole interaction terms in the Hamiltonian. The enhanced $E1$ transitions between the two parity states, as obtained in the present experimental work on  $^{100}$Ru, are considered an indication of the breakdown of the reflection symmetry. The obvious method to consider the reflection asymmetry is to include odd-multipole interaction terms in the Hamiltonian and mix both parity states in the self-consistent mean-field calculations. In a few regions of the nuclear chart, the mean-field calculations lead to a minimum in the potential energy surface as a function of the octupole deformation parameter, $\beta_3$. There have been a number of studies using different approaches to investigate the importance of reflection asymmetric deformation and several review articles have been published \cite{butler96,butler16}. However, in all these studies octupole deformation has been discussed for the ground-state
configuration only. For $^{100}$Ru, the negative parity band is a two-quasiparticle band and consequently, multi-quasiparticle states need to be included in the basis space in order to investigate octupole correlations for this system.
\begin{figure}[!ht]
    \begin{center}
    \vspace{3.1cm}
       \hspace*{0cm}\includegraphics[height=5.5cm, width=7.5cm, angle =0]{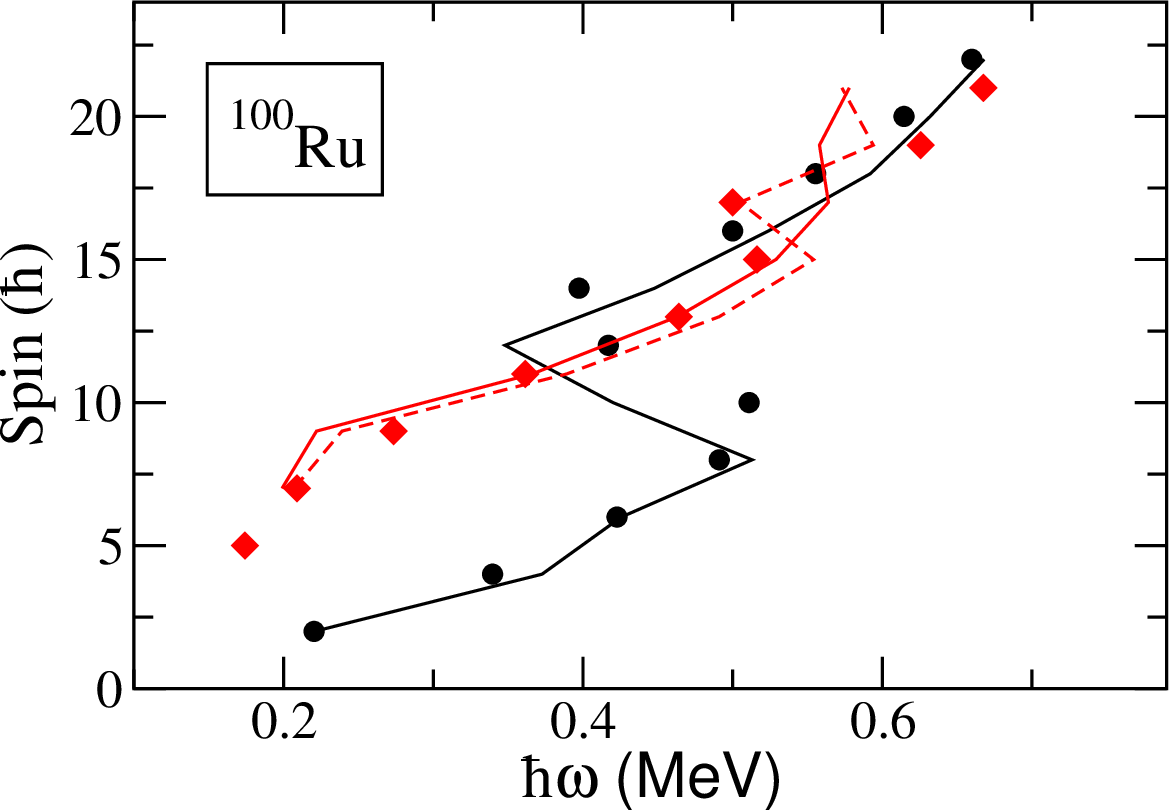}
    \end{center}
    \caption{ (Color online) Level energies
      as a function of spin and rotational frequency ($\hbar $$\omega$),  for yrast $+$ve parity band (black circles) and yrast $-$ve parity band (red diamond). The solid line represents the values from TPSM calculations and the dotted line represents the TPSM values with reduced pairing.}
    \label{energy_rotor}
\end{figure}
As a first step to consider octupole correlations in the multi-quasiparticle TPSM approach, we have augmented the Hamiltonian by including octupole-octupole interaction terms among neutrons, protons, and also between neutrons and protons. The coupling constant of octupole interaction term is adopted from the hydrodynamical estimate \cite{BM} as has been done in the Hartree-Fock-Bogoliubov study \cite{ring}. The Hamiltonian is diagonalized with the angular-momentum projected wavefunction having parity as a good quantum number and, therefore, in the present work, octupole interaction is considered as a perturbation. The details of the extended version of the TPSM approach shall be presented in a forthcoming publication \cite{naz24}.
  
The TPSM calculations for both $+$ve and $-$ve parity states have been performed with the axial and nonaxial deformation parameters of $\epsilon = 0.24$ and $\epsilon' = 0.14$. The pairing parameters are $\Delta_{n}= 1.08$ MeV and $\Delta_{p}= 1.03$ MeV. It is quite possible that mean-field parameters will be different for the $-$ve parity states as the low-lying states are two-quasiparticle states. However, in the absence of any systematic study, we have adopted the same mean-field parameters for both parities. Furthermore, in the TPSM approach, the multi-quasiparticle states form the basis space, and the diagonalization of the shell model Hamiltonian within this space should subsume minor modifications in the mean-field. The basis space in the TPSM analysis consists of three major oscillator shells for neutrons ($N = 3, 4, 5$) and three for protons ($N = 2, 3, 4$). For the $+$ve parity states, the excitations are considered from the last oscillator shell, and for the $-$ve parity states, the excitations are included from the last two oscillator shells \cite{NAZ_23} which have different parities. The vacuum configurations are generated with all the three major shells \cite{JAS99}.\par

\begin{figure}[!htb]
    \begin{center}
    \vspace{1cm}
       \includegraphics[height=5cm, width=7.5cm, angle =0]{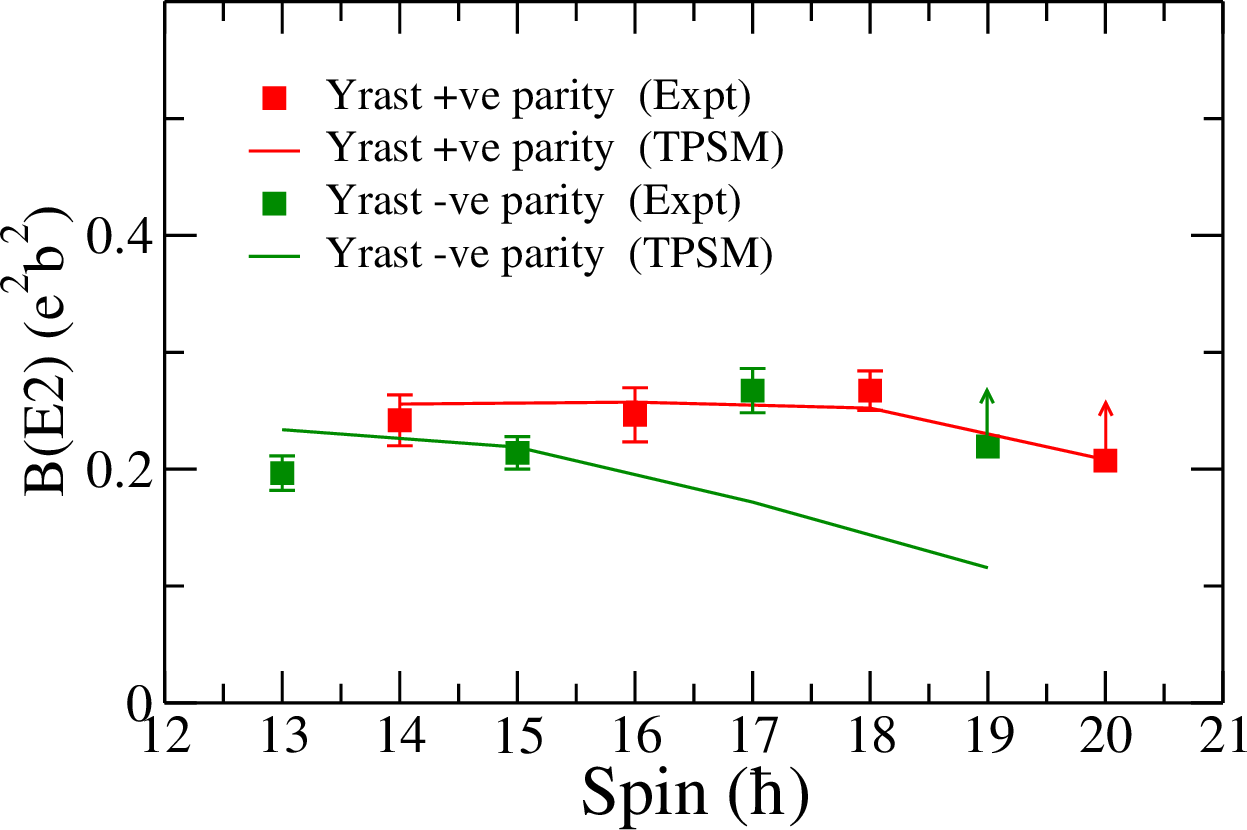}
    \end{center}
    \caption{(Color online) The in-band $B(E2)$ transition rates as a function of spin for yrast $+$ve parity band (Band 2) and yrast $-$ve parity band (Band 3) in $^{100}$Ru. The solid line represents the values from TPSM calculations.}
    \label{be2_100ru}
\end{figure}

The calculated TPSM energies after configuration mixing are displayed in Fig.~\ref{level_energy} along with the experimental data for the lowest $+$ve and $-$ve parity bands. It is evident from the figure that TPSM reproduces the observed energies remarkably well for both the parity bands. There are small deviations for the highest observed spin states and is about 0.29 MeV for $I = 26\hbar$. These deviations are expected as for such a high-spin state, multi-quasiparticle states of four neutrons and four protons become important, which have not been considered in the present work. We have also plotted angular momentum versus rotational frequency in Fig.~\ref{energy_rotor} in order to explore the band-crossing phenomena.  It is noted from the figure that band-crossing frequency at around $\hbar \omega = 0.4$ MeV for the $+$ve parity band is reasonably well reproduced by the TPSM calculations. For the $-$ve parity band, the band crossing at $\hbar \omega = 0.5$ MeV is over-predicted by the TPSM approach with the same parameter set as for the $+$ve parity band. We have also performed calculations with the reduced pairing for the $-$ve parity band as this is a two-quasiparticle band and the pairing is expected to be reduced. The calculations with a reduced pairing of $\Delta_n = 0.72$ MeV and $\Delta_p = 0.85$ MeV are shown by the dotted line in Fig.~\ref{energy_rotor}. It is evident that band-crossing frequency is now reproduced quite well. However, the discrepancies for the spin states of $I = 19\hbar$ and $21\hbar$ remain even with the reduced pairing. The band-crossing in both bands is due to the alignment of two neutrons and can be easily delineated from the analysis of the TPSM wavefunctions.

We have evaluated the transition probabilities using the TPSM wavefunctions with the expressions provided in previous publications \cite{SUN02, EPJ21, spr24}. For the electric transitions, we have used the effective charges of $e_p = 1.5 e$ and $e_n = 0.5 e$ for protons and neutrons, respectively. The $B(E2)$ transitions for the $+$ve parity yrast band and $-$ve parity yrast band are plotted in Fig.~\ref{be2_100ru}. The observed $B(E2)$ transitions for the $+$ve parity band are reproduced quite well by the TPSM approach, however, for the $-$ve parity large deviations are noted for spin states of $I = 17\hbar$ and $19\hbar$. The reason for this deviation is not clear at this stage and more studies are needed using other theoretical models.

\begin{figure}[!ht]
    \begin{center}
       \vspace*{0.5cm}\includegraphics[height=5cm, width=7.5cm, angle =0]{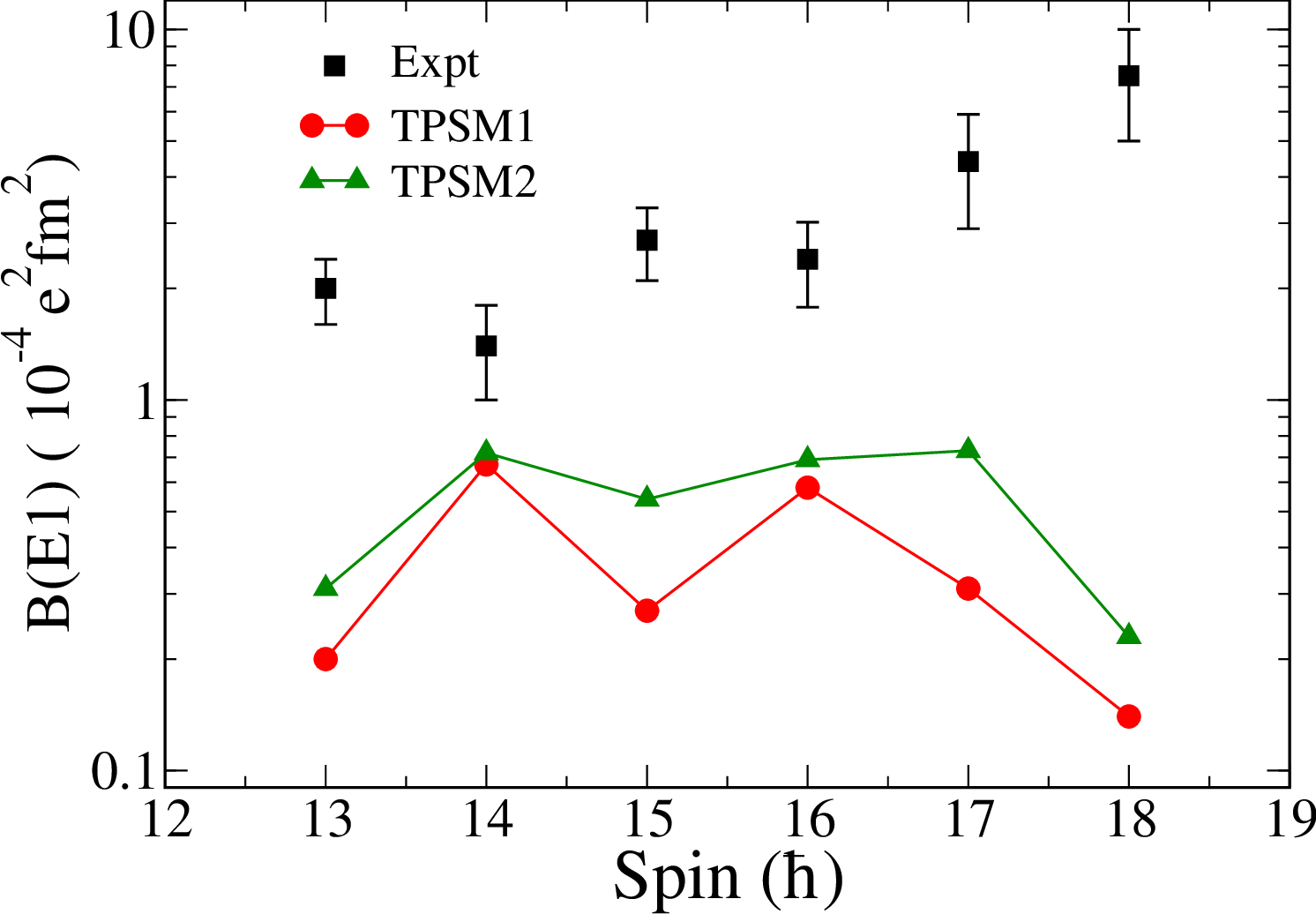}
    \end{center}
    \caption{(Color online) $B(E1)$ values vs spin for Band 2 and Band 3 in $^{100}$Ru. The solid lines represent theoretical calculations using the TPSM framework. TPSM1 (TPSM2) is without (with) octupole interaction.}
    \label{be1_100ru}
\end{figure}

We shall now turn our discussion to the $B(E1)$ transitions which form a crucial element in determining whether the system has octupole deformation. It is known that for systems with octupole collectivity, these transitions are strongly enhanced \cite{ba2}. In an attempt to address this question, we have calculated $B(E1)$ transitions with and without octupole-octupole interaction in the Hamiltonian, and the results are presented in Fig.~\ref{be1_100ru}.
In the absence of octupole-octupole interaction, the transitions are noted to be an order of magnitude weaker compared to the observed values. The inclusion of octupole-octupole interaction raises the magnitude of the transition rates and for some transitions by a factor of two, but are still quite weak in comparison to the observed rates. The reason for obtaining weak $B(E1)$ transitions is that in the present work, we have included octupole-octupole interaction as a perturbation with the mean-field containing only the quadrupole degree of freedom and the parity remains conserved in the calculations at all levels. Thus, it can be
stated from the comparison in Fig.~\ref{be1_100ru} that it is necessary to consider octupole collectivity for $^{100}$Ru, and the mean-field Hamiltonian should break the reflection symmetry.\par

It may also be noted that the dynamic octupole deformation (vibrational octupole mode) in $^{100}$Ru may be
ruled out from the following arguments:\\
1. For a band built on a rotationally aligned octupole phonon, the negative parity sequence is always shifted up in energy \cite{hageman}, and thus, the two opposite parity bands are not interspaced.\\
2. Only the transitions I$^-$ $\xrightarrow{}$ (I $-$ 1)$^+$ are allowed \cite{frau_prc}. Thus, the interleaved $E1$ transitions with nearly constant transition rates, will not be observed.

To summarize, the level lifetimes of the alternate parity bands of $^{100}$Ru have been measured. The estimated $B(E1)$ rates for the seven interleaved $E1$ transitions were found to be enhanced by an order of magnitude compared to those calculated using the TPSM approach with the inclusion of octupole-octupole interaction, but preserving the reflection symmetry. This comparison suggests the presence of stable octupole deformation in $^{100}$Ru based on a two-quasiparticle configuration. More detailed calculations with broken reflection symmetry in the TPSM approach and other theoretical models are needed to confirm the octupole nature of the bands at high spin and to rule out the mundane two-quasiparticle interpretation of the band structures.

\begin{acknowledgments}
The authors would like to thank the operational staff of the K-130 cyclotron at VECC for providing a good quality beam as well as necessary support during the pandemic period. The authors are thankful to the Department of Atomic Energy and the Department of Science and Technology, Government of India for providing the necessary funding for the clover array. A.K. acknowledges the grant from the Council of Scientific \& Industrial  Research (CSIR) (File No: 09/489(0121)/2019-EMR-I), Government of India. N.N. acknowledges the grant from the DST-INSPIRE Fellowship (No. DST/INSPIRE Fellowship/2020/IF200508 ). 
\end{acknowledgments}

\bibliographystyle{apsrev4-1}

\end{document}